\documentclass[conference]{IEEEtran}
\usepackage[english]{babel}

\usepackage{cite}
\usepackage{amsmath,amssymb,amsfonts,amsthm,amssymb}
\usepackage{algorithmic}
\usepackage{float}
\usepackage{graphicx}
\usepackage{subfig}
\usepackage{textcomp}
\usepackage{breqn}
\usepackage{color}
\usepackage[colorinlistoftodos]{todonotes}
\usepackage[colorlinks=true, allcolors=blue]{hyperref}

\usepackage{array,multirow,makecell}
\setcellgapes{1pt}
\newcolumntype{R}[1]{>{\raggedleft\arraybackslash }b{#1}}
\newcolumntype{L}[1]{>{\raggedright\arraybackslash }b{#1}}
\newcolumntype{C}[1]{>{\centering\arraybackslash }b{#1}}

\IEEEoverridecommandlockouts 

\begin{document}

\title{Driving Path Stability in VANETs}

\author{
	\IEEEauthorblockN{ 
		Mohammed Laroui\IEEEauthorrefmark{1}\IEEEauthorrefmark{3},
		Akrem Sellami\IEEEauthorrefmark{3},
        Boubakr Nour\IEEEauthorrefmark{4},
		Hassine Moungla\IEEEauthorrefmark{3}\IEEEauthorrefmark{5},
        Hossam Afifi\IEEEauthorrefmark{5},
		and Sofiane B. Hacene\IEEEauthorrefmark{1}
	}

	\IEEEauthorblockA{\IEEEauthorrefmark{1}Evolutionary Engineering \& Distributed Information Systems Laboratory, Djillali Liabes University, Sidi Bel Abbes, Algeria}
	\IEEEauthorblockA{\IEEEauthorrefmark{3}LIPADE, Paris Descartes University, Sorbonne Paris Cite, Paris, France}
    \IEEEauthorblockA{\IEEEauthorrefmark{4}School of Computer Science, Beijing Institute of Technology, Beijing, China}
	\IEEEauthorblockA{\IEEEauthorrefmark{5}UMR 5157, CNRS, Institute Mines Telecom, Telecom SudParis, Nano-Innov CEA Saclay, France}
	Emails: \{mohammed.laroui,boukli\}@univ-sba.dz, n.boubakr@bit.edu.cn\\ \{mohammed.laroui,akrem.sellami,hassine.moungla\}@parisdescartes.fr, hossam.afifi@telecom-sudparis.eu
	
	\thanks{Mr. Laroui is corresponding author.}
}

\maketitle

\begin{abstract}
    Vehicular Ad Hoc Network has attracted both research and industrial community due to its benefits in facilitating human life and enhancing the security and comfort. However, various issues have been faced in such networks such as information security, routing reliability, dynamic high mobility of vehicles, that influence the stability of communication. To overcome this issue, it is necessary to increase the routing protocols performances, by keeping only the stable path during the communication. The effective solutions that have been investigated in the literature are based on the link prediction to avoid broken links. In this paper, we propose a new solution based on machine learning concept for link prediction, using LR and Support Vector Regression (SVR) which is a variant of the Support Vector Machine (SVM) algorithm. SVR allows predicting the movements of the vehicles in the network which gives us a decision for the link state at a future time.  We study the performance of SVR by comparing the generated prediction values against real movement traces of different vehicles in various mobility scenarios, and to show the effectiveness of the proposed method, we calculate the error rate. Finally, we compare this new SVR method with Lagrange interpolation solution.
\end{abstract}


\section{Introduction}

In Wireless Networks \cite{WirelessNetwork,moungla2016distributed,moungla2014cost}, the communication is based radio waves (radio and infrared) instead of the usual wired cables. This technology supports users high mobility, and allows them to communicate with each other during movement without requiring to use of cables. Thus, this kind of communication is a big challenge for future generation of network, especially to apply new technologies of communication.

Mobile Wireless Networks\cite{MobileWirelessNetwork} allow an easy connection of nodes between 10 meters and a few kilometers, it can be classified into two classes: (a) Networks with Infrastructure (cellular networks), and Networks without Infrastructure (Adhoc networks) \cite{Adhoc_and_cellular_network}. In the cellular model, the communication between the nodes is managed by the base stations. Whereas in ad hoc networks, the communication is achieved without any infrastructure. A wireless vehicular network \cite{Vanet} is an ad-hoc network, where vehicle nodes circulate in a road, city, etc, it uses multi-jump communications with specific routing protocols \cite{moungla2012reliable}.

Various routing protocols that can be used in wireless networks are proposed in the literature \cite{Routing_Protocol}.  VANET-based solutions are expected to furnish methodical and proven solutions for innovative, and resource-efficient as in other wireless communication protocol \cite{CCNC}. These protocols can broadly be classified into three main categories: the first category is the proactive routing protocols, which aim to construct the routing tables before the request is made. A proactive routing protocol identifies the topology of the network at all times, for example, Destination Sequenced Distance Vector routing (DSDV). The second category is the reactive routing protocols, that consist of building a routing table only when a node receives a request. Protocols under this umbrella do not know the network topology; they determine the path to access a node of the network due to the demand of request, for example, Ad hoc On-Demand Distance Vector (AODV). Finally, the last category is the hybrid routing protocols. A hybrid protocol combines the two previously discussed categories: proactive and reactive concept. It uses a proactive protocol to get information about the nearest neighbors (maximum neighbors with two jumps). Beyond this predefined area, the hybrid protocol uses reactive protocol techniques to search for routes. This type of protocol adapts well to large networks.

The main characteristics of VANET networks are the high mobility of vehicles, where each vehicle has a range of communication to provide communication directly with the destination vehicle if they are in the same range. Otherwise, a multi-hop communication needs to be established to allow communication with the destination vehicle. During the communication, the location of vehicles change rapidly because of the high mobility of vehicles (moving with high speed in a short time) this will affect the stability of path during the communication with the risk of breaking the communication path. This issue is considered as one of the main problems in VANET due to its consequence in the quality of service \cite{mekki2017vehicular}.

The main motivation of this work is to study the path communication stability in VANETs. Thus, we propose in this work a mechanism based on LR and SVR to guarantee the path stability and enhance the quality of service within VANET applications.

The remaining of this paper is organized as follows: in the following section, we present related works. In section III, we describe our model of prediction (SVR). In Section IV, the experimental results are presented. And finally; we conclude the paper in Section V.

\section{Related Work}
In this section, we present the state of art solutions proposed to resolve the problem of link failure in VANET networks.

Crisòstomo et al. \cite{Crisostomo} proposed a local route repair, nodes in the network trigger the repair procedure When they predict that a link on the route to a destination is about to break. The work assumes that all nodes in the network are equipped with GPS. Also, all packets are modified, to contain node positions and motion information obtained using GPS. The major limitation of this approach is the cost of using GPS and resource waste due to synchronization, and still, the GPS does not detect the obstacles.

Work in \cite{Boukerche} proposes a Preemptive AODV protocol (PrAODV). Authors tried to combine \textit{Schedule a Rediscovery in Advance}, and \textit{Warn the Source Before the Path Breaks} mechanisms. The latter mechanism is attained by exchanging ping–pong Message (Hello message). The comparison is made between the signal power of received packets and a threshold value. When the power of the packet signal is less than the threshold value, the node sends its neighbors a message called "Ping", the neighboring nodes must respond with a message called "Pong", within a fixed timeout if the node does not receive this message, a warning message should be returned to the source node that begins a rediscovery. When the source receives a response packet from an intermediate node, it collects the link information. Therefore, when the source node receives the packet containing the status information of all links, including the minimum TTL value of the links. Thus, we can schedule a rediscovery time before the path breaks. The major problem with this technique is the high routing cost.

Similarly, work in \cite{Boukli} proposes a predictive and preemptive Maintenance of routes (PPAODV), aiming to improve the performance (quality of service) of the Ad-hoc On-Demand Distance Vector (AODV). PPAODV allows predicting the movements of the nodes using the history (last movements). The protocol considers that a node is in a dangerous or preemptive zone if it receives a data packet from a predecessor node with the signal strength below a signal level $P$. The prediction of failure of The link is calculated by Lagrange interpolation \cite{Boukli}. When two consecutive measurements give the same signal strength, the time of the second occurrence is stored. The expected signal strength $P$ of the received packets from the predecessor node is calculated as follows:

\begin{align}
P=\bigl(\begin{smallmatrix}
\frac{\left ( t-t_{1} \right )\times \left ( t-t_{2} \right )}{\left ( t_{0}-t_{1} \right )\times \left ( t_{0}-t_{2} \right )}\times P_{0}
\end{smallmatrix}\bigr)+
\bigl(\begin{smallmatrix}
\frac{\left ( t-t_{0} \right )\times \left ( t-t_{2} \right )}{\left ( t_{1}-t_{0} \right )\times \left ( t_{1}-t_{2} \right )}\times P_{1}
\end{smallmatrix}\bigr)+ ~\\ \nonumber
\bigl(\begin{smallmatrix}
\frac{\left ( t-t_{0} \right )\times \left ( t-t_{1} \right )}{\left ( t_{2}-t_{0} \right )\times \left ( t_{2}-t_{1} \right )}\times P_{2}
\end{smallmatrix}\bigr)
\end{align}
With : 
\begin{align}
t=t_{2}+\bigl(\begin{smallmatrix}
\frac{t_{0}+t_{1}+t_{2}}{3}
\end{smallmatrix}\bigr)+ Discovery~period
\end{align}

Hayato Kitamoto et al. \cite{kitamoto} proposed a High Precision-PPAODV (HP-PPAODV), which is an improvement of the prediction defined in \cite{Boukli}, and allows the use of the interpolation of Newton, adding the number of RSS acquisition and predicting the value of TDP. The accurate prediction of link failure can improve the packet delivery ratio. The Newton interpolation for RSS
computation is shown as:

\begin{align}
P(t_{PT})=P^\prime + \displaystyle\prod_{i=1}^{2} (t_{PT} - t_{i})f [\! t_{1},t_{2},t_{3} \!]
\end{align}
Where :
\begin{align}
f [\! t_{1},t_{2},t_{3} \!] =\frac{f [\! t_{2},t_{3} \!]-f [\! t_{1},t_{2}\!]}{t_{3}-t_{1}}
~\\
f [\! t_{1},t_{2} \!] =\frac{f [\! t_{2}\!]-f [\! t_{1}\!]}{t_{2}-t_{1}}
~\\
f [\! t_{1}\!]=P_{1} , f [\! t_{2}\!]=P_{2} , f [\!t_{3}\!]=P_{3}
\end{align}
$P^\prime$ represents the previous predicted RSS.
By using Discovery Period $T_{DP}$, Predict Time $(t_{PT})$ is
shown as:
\begin{align}
t_{PT}=t_{3}+T_{DP}
\end{align}

Work in \cite{Vinod} proposes a prediction based routing protocol (PBR) that predicts the life of each route and creates new routes to replace old ones before they break. Authors used a Highway Mobility Model because highway scenario is characterized by high speeds of vehicles. In order to characterize the movement of vehicles on the road they used either the macroscopic (traffic density, traffic flow) or microscopic approach for objective to generate vehicle movement patterns for wireless ad hoc routing.Two important pieces of information in PBR protocol are location and velocity information of vehicles on the route to the gateway, and the prediction algorithm uses this information to predict when the route will break. It is unclear how a vehicle shares its wireless Internet connection with others vehicles.

Authors in \cite{shelly} studied the link life-time in VANETs using analytical model for Probability Density Function (PDF), they analyzed the statistics of link life time and the impact of transmission range, vehicle mobility, and vehicle density; by considering the scenario of communication is un-congested free flow traffic, where the state of  movement of vehicles is independent. Authors were interested in the movement of two vehicles (A) and (B), with $V_{A},V_{B}$ and $V_{r}$ are  velocities of vehicle A, velocities of vehicle B respectively, and the relative velocities between pair of vehicles in the network $V_{r}=V_{A}-V_{B}, V{r}$ is in $(-\upsilon_{m},+\upsilon_{m})$ Where $\upsilon_{m}=\upsilon_{max}-\upsilon_{min}$. In this solution, they supposed that the velocity of vehicles follows a uniform distribution.

Work in \cite{Zhang2016} proposes a link duration prediction via AdaBoost algorithm \cite{AdaBoost}. The proposed steps consist of aggregate the existing link metrics to generate many predictors, each predictor $f(x,d)$ predicts if the link duration is under or over $d$ with high accuracy using the set of link metrics $x$. In the next step, the algorithm determines the duration of the link using all the knowledge collected from these predictors. Authors considered that two vehicle can communicate with each other if the distance between them is less than $R$ meters where $R$ is consider less than the transmission range. Each vehicle $v$ entries their neighbors $u$. When the distance to appear is less than $R$, the vehicle $v$ waits a random time, and then calculates the link metric $x$ with the vehicles $u$, and inserts them into the entry. When the distance between $u$ and $v$ is at least $R$ meters, the last updating of link metrics is inserted into the entry. $M= \left\lbrace m_{1},m_{2},...,m_{r}\right\rbrace $ is a set of available link, and $L(e)$ is uploaded from RSU. all vehicles can upload all messages about its expired links with all or part of their neighbors. From AdaBoost algorithm, they construct strong classifiers from weak classifiers. The classifier corresponding for each link duration $jD$ is obtained by AdaBoost algorithm. Based on locally link metrics, each vehicle can predict the link duration. The limitation of AdaBoost algorithm is when he is applied in multi-classes problems.

Authors in \cite{ELDP} proposed an extended link duration prediction (ELDP) model that allows vehicles in the network to estimate the duration of the connection to other vehicles.
The simulations of scenarios in highway and city illustrate that the proportional speed between vehicles has an impact on the prediction of link duration in VANET. Also, the turning directions of vehicle in crossroad have a direct impact on the prediction results. The limit of this work, it needs to suppose that vehicles speed must follow a normal distribution.

Work in \cite{vanet2017new} proposes NGOMA algorithm (Network formation Game for MAC-level retransmission) for cooperative communication in VANET. The proposed algorithm selects a node from the set of intermediate nodes in the MAC-layer communication process that incorporates the store-and-forward process. In case of network failure, the best relay node is selected from a set of neighboring nodes using the network formation game technique in order to retransmit a packet from the source to destination. NGOMA protocol enhances packet delivery ratio and reduces delay. This work focuses only on allocating the resource, for example, bandwidth sharing and channel access.

Similarly, work in \cite{Predicting_Using_LED} predict link failure in a route using data forwarding technique by generating a link existence diagram (LED) that allows knowing the existing link between vehicles, this technique reduces the end-to-end delay. However, this approach uses GPS that may not detect obstacles and need of resources.

Each solution proposed by the researchers has its advantages and disadvantages. However, most of them tried to solve the problem in order to improve the quality of service. Our main contribution is to propose link failure prediction methods LR and SVR \cite{SVR} that allows vehicles to check their status in the future to prevent link failure in VANETs.

\section{SVR-based of-Path Stability}
Support Vector Regression (SVR) \cite{SVR} is a machine learning algorithm used to perform prediction based on historical data as a training set. It uses minimization of structural risk (SRM) instead of minimizing empirical risk (ERM).

The function is formulated as follows:
\begin{equation}
f(x)=W^{T}\phi(x)+b 
\end{equation}

\textit{W} is a vector in $\textbf{F}$ space and $\phi$ is the transformation function corresponding to $\textbf{F}$. The function is obtained by solving the following primal problem: 
\begin{equation}
\displaystyle\mathop{min}_{w, b} \ {\frac{1}{2}}\Vert w\Vert^{2}+C\sum\limits_{i=1}^{N}(\xi_{i}+\xi_{i}^{\ast}). 
\end{equation} 
\begin{align*}
s.t.\ & ((w\bullet x_{i})+b)-y_{i}\leq\varepsilon+\xi_{i},i= 1,2,\ldots, N \\ & y_{i}-((w\bullet x_{i})+b)\leq\varepsilon+\xi_{i},i=1,2,\ldots, N \\ 
& \xi_{i}^{\ast}\geq 0,i= 1,2,\ldots, N
\end{align*}

Where $\xi_{i}$ and $\xi_{i}^{\ast}$ are slack variables introduced to deal with prediction errors higher than the insensitive loss parameter $\epsilon$ and C is the penalty parameter.

To solve the quadratic programming of the primal formulation, Lagrange multipliers are introduced, and the following dual formulation is obtained: 
\begin{equation}
\begin{aligned}
\displaystyle\mathop{\min}_{\alpha,\alpha^{i}} \  {\frac{1}{2}}\sum\limits_{i}^{N}\sum\limits_{j}^{N}K(x_{i},x_{j})(\alpha_{i}-\alpha_{i}^{\ast})(\alpha_{j}-\alpha_{j}^{\ast})+\\
\varepsilon\sum\limits_{i}^{N}(\alpha_{i}+\alpha_{i}^{\ast})-\sum\limits_{i}^{N}y_{i}(\alpha_{i}-\alpha_{i}^{\ast})
\end{aligned}
\end{equation}
\begin{align*}
s.t.\ & \sum\limits_{i=1}^{N}(\alpha_{i}-\alpha_{i}^{\ast})=0, \cr & 0\leq\alpha_{i},\alpha_{i}^{\ast}\leq{\frac{c}{N}},i=1,2,\ldots,N
\end{align*}

The optimal prediction function is found after introducing Lagrangian multipliers ($\alpha_{i}$) and it is as follows: 
\begin{equation}
f(x)=\sum\limits_{i=1}^{l}(\alpha_{i}-\alpha_{i}^{\ast})K(x_{i},x)+b
\end{equation}
where $K(x_{i},x_{j}) = \phi(x_{i})^{T}\phi(x_{j})$ is the Kernel function. In our contribution, we use the radial basis function (RBF), $K(x_{i},x_{j}) = \exp(-\gamma\Vert x_{i}-x_{j}\Vert)^{2}$ as kernel function because it is more efficient with non-linear time-series (as our case) due to its generalization ability and non-linear mapping ability into a infinite feature space.
~\\

\subsection{Prediction Of The Vehicular Trajectory}
We propose in this work an algorithm based on SVR to predict the next positions for each vehicle for different mobility use cases.~\\
As a supervised computer learning, SVR must be trained at first to build the prediction model. For these needs, we extract a set of several vehicle trajectories from SUMO, and for three different use cases, such as a congested scenario in a city center, a road scenario and a part of an intersection.
~\\
Since a vehicular trajectory can not carry on cyclical grounds and that the trajectory can be presented with many stochastic models because we can observe several deviations and sudden events in a trajectory, we propose to use four models SVR, adjusted beforehand with the best combination of parameters (according to the trend of the trajectory and its speed history) to improve the prediction results of the trajectory:
~\\
-SVR1: Used when vehicle speed is high and constant.~\\
-SVR2: Used if the vehicle speed decreases slowly, then it increases (case of a turn).~\\
-SVR3: Used if the vehicle speed is very low and decreases until stop. (Parking or intersection).~\\
-SVR4: Used if the vehicle was stopped and starts to move. (Increasing speed).
~\\
The choice of the SVR model at instant \textit{t} is based on the analysis of the previous \textit{t-k} values of the vehicle speed and the distribution of the \textit{t-k} previous geographical positions. Depending on this, the chosen model is trained with the data of the previous k positions to predict future positions and so on.

\section{Experimental results}
\begin{figure}[!t]
	\centering
	\includegraphics[width=3in]{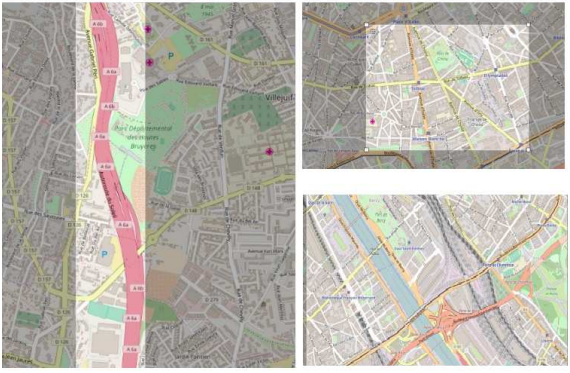}
	\caption{\label{fig:all_scenario}Simulation Scenarios}
\end{figure}

This section aims to evaluate the proposed prediction method. Therefore, we study different use cases in the simulation: city center, a road scenario (sun highway in France), and a part of an intersection, Figure \ref{fig:all_scenario}. 

\subsection{Scenarios description}
We generated the three use cases from OpenStreetMap website~\footnote{OpenStreetMap: www.openstreetmap.org}. After this step, we generated a mobility file for each use case in order to apply it in SUMO~\footnote{SUMO: www.sumo.dlr.de} (Simulation for Urbain Mobility). Therefore, for each use case, we study the prediction generated by LR and SVR methods, and compare the results with the prediction generated with Lagrange interpolation. Finally, the three results are compared with the real movement of the vehicle.

\begin{figure*}[!t]
	\centering
	\subfloat[Road Scenario]{
		{\includegraphics[width=0.32\linewidth]{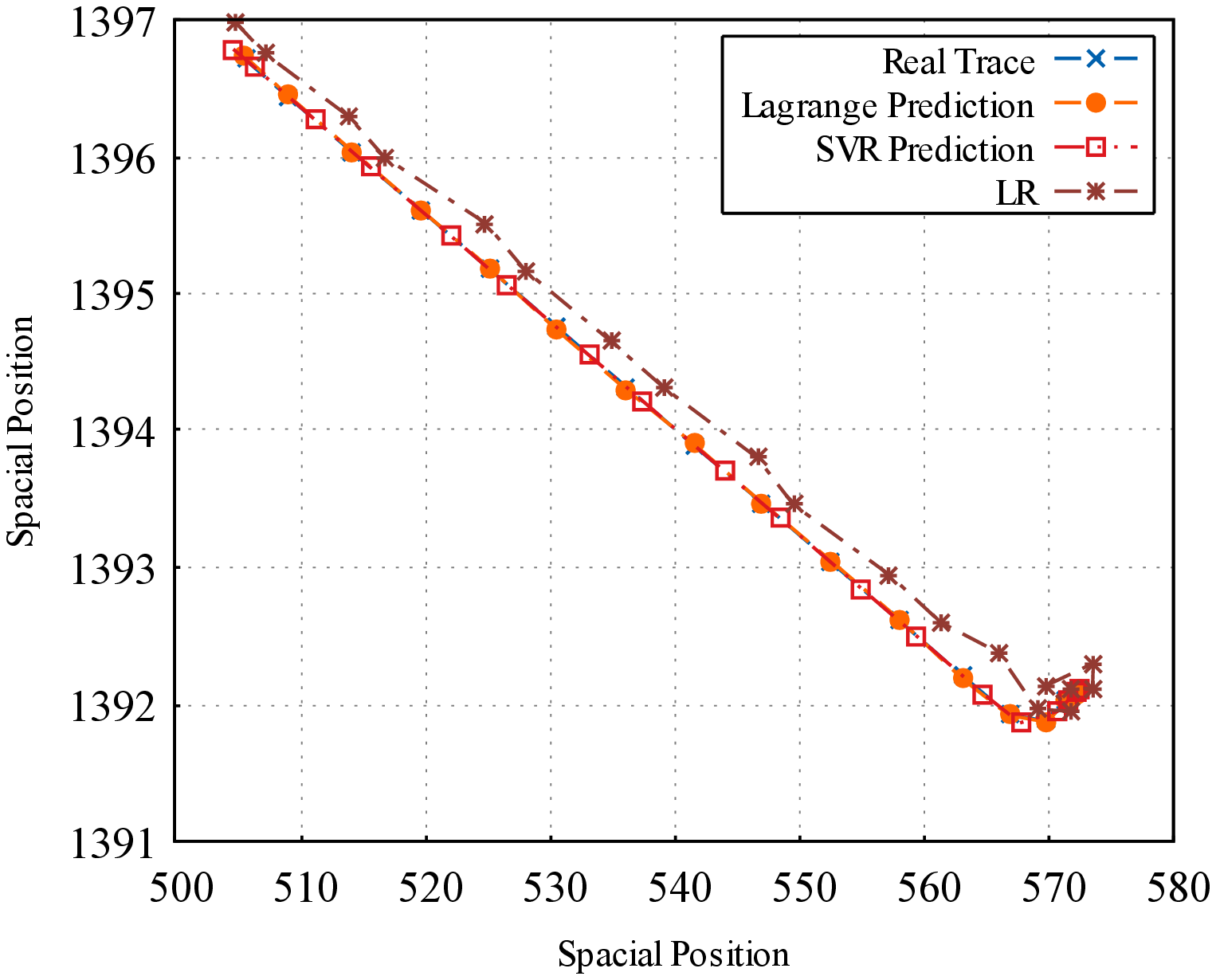}}
		\label{fig:predicted_road}
	}
	\subfloat[Intersection Scenario]{
		{\includegraphics[width=0.32\linewidth]{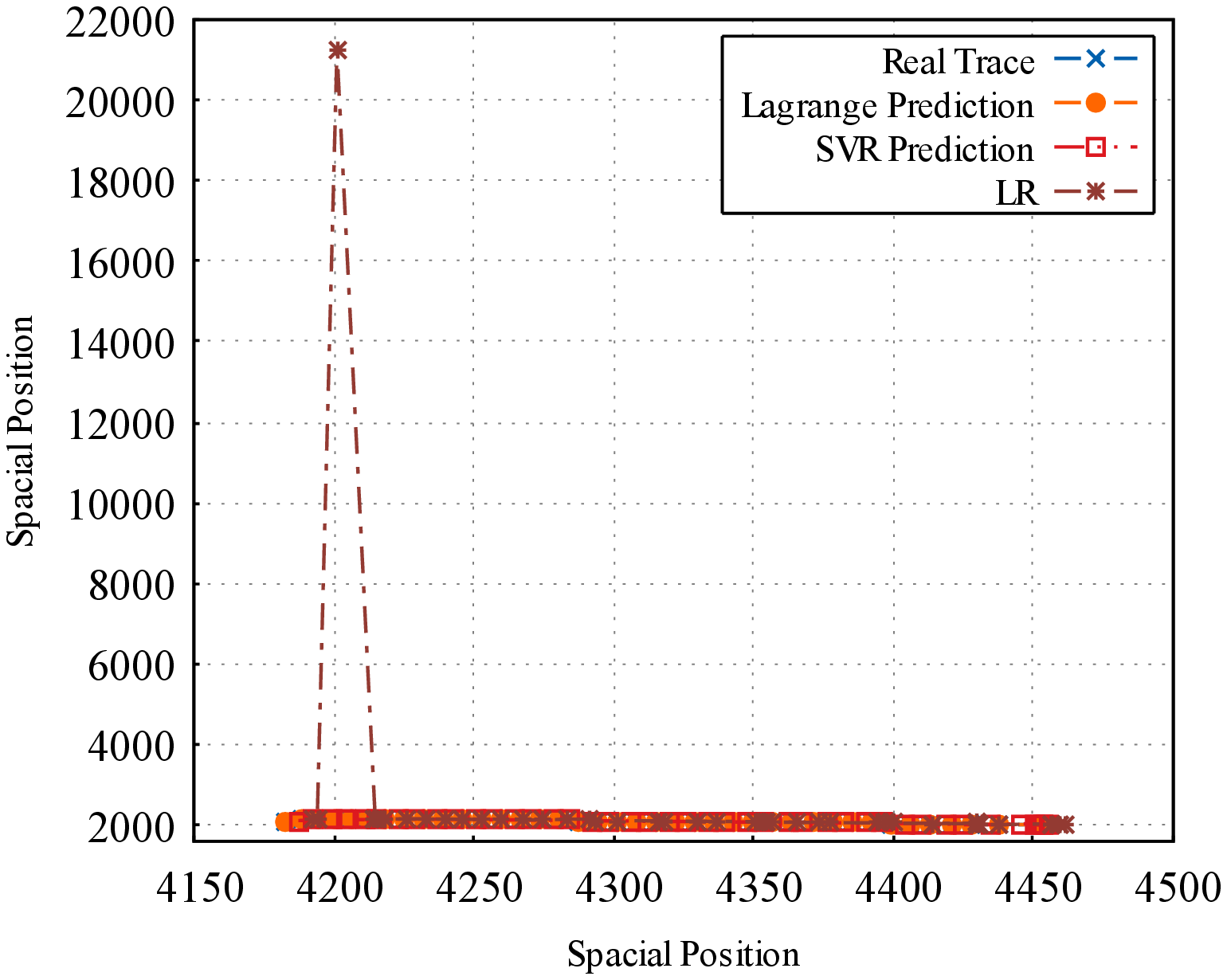}}
		\label{fig:predicted_intersection}		
	}
	\subfloat[City Scenario]{
		{\includegraphics[width=0.32\linewidth]{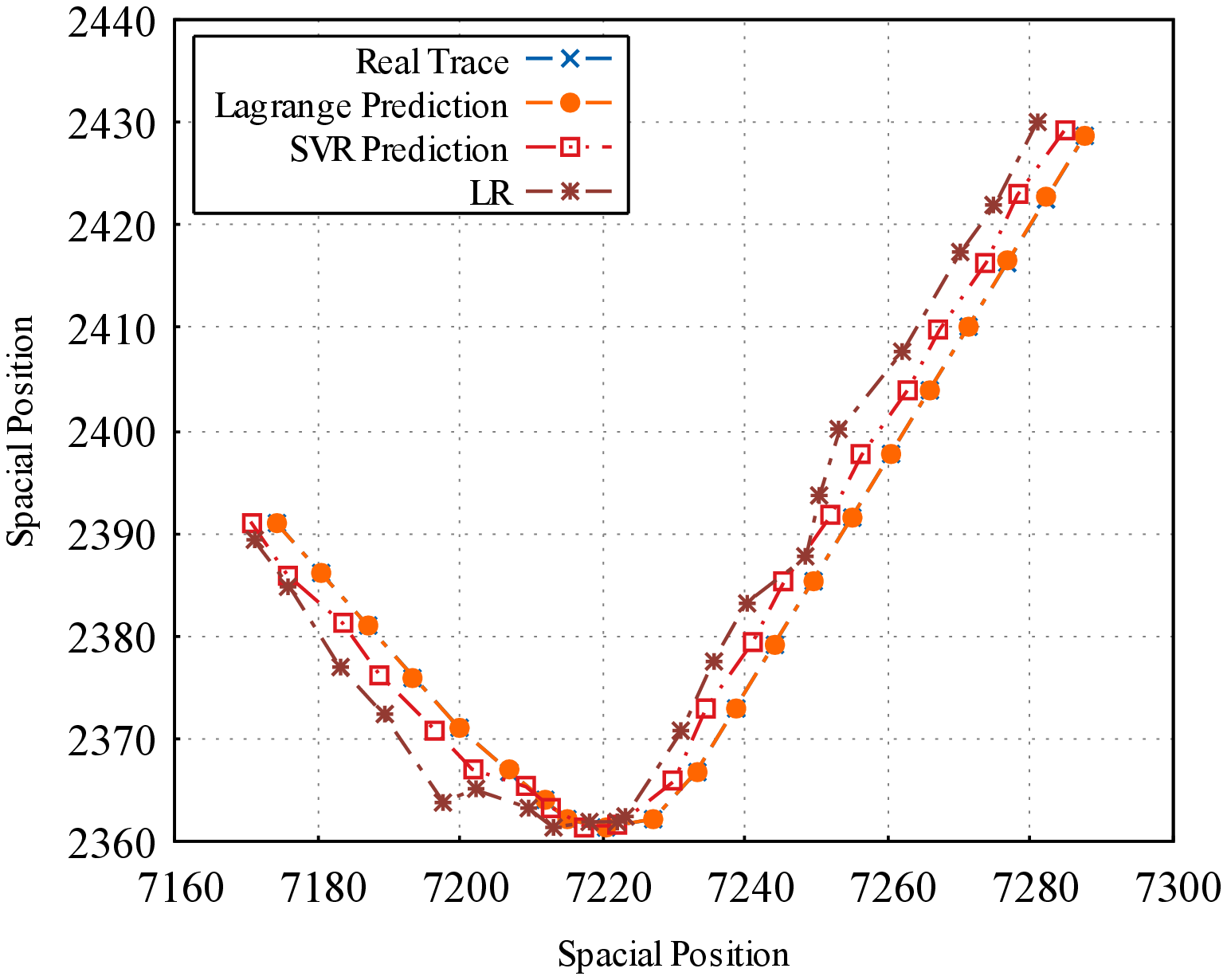}}
		\label{fig:predicted_city}		
	}
	\caption{Predicted traces using Lagrange, SVR, and LR }
	\label{fig:performance_evaluation}
\end{figure*}

Figure \ref{fig:predicted_road} represents the movement of vehicle N°5 in the road scenario, all plots in the figure are displaying the real movement of the vehicle and the predicted movement using Lagrange, LR, and SVR methods. We can see that the three methods gave results close to the real movement in the bend of the road. However, Lagrange is more efficient compared to LR and SVR in the highway.

Figure \ref{fig:predicted_intersection} depicts the movement of vehicle N°14 in intersection scenario. The predicted traces of Lagrange and SVR are more close to real traces compared to LR with a slight preference for Lagrange method.

Figure \ref{fig:predicted_city} represents the movement of vehicle N°7 in city scenario. The predicted traces with Lagrange are close to real movement in compared to LR and SVR methods. However, when the vehicle moves in the bend, the three prediction methods are close to real movement.

\subsection{Performance Analysis}
\begin{figure*}[!t]
	\centering
	\subfloat[Road Scenario]{
		{\includegraphics[width=0.31\linewidth]{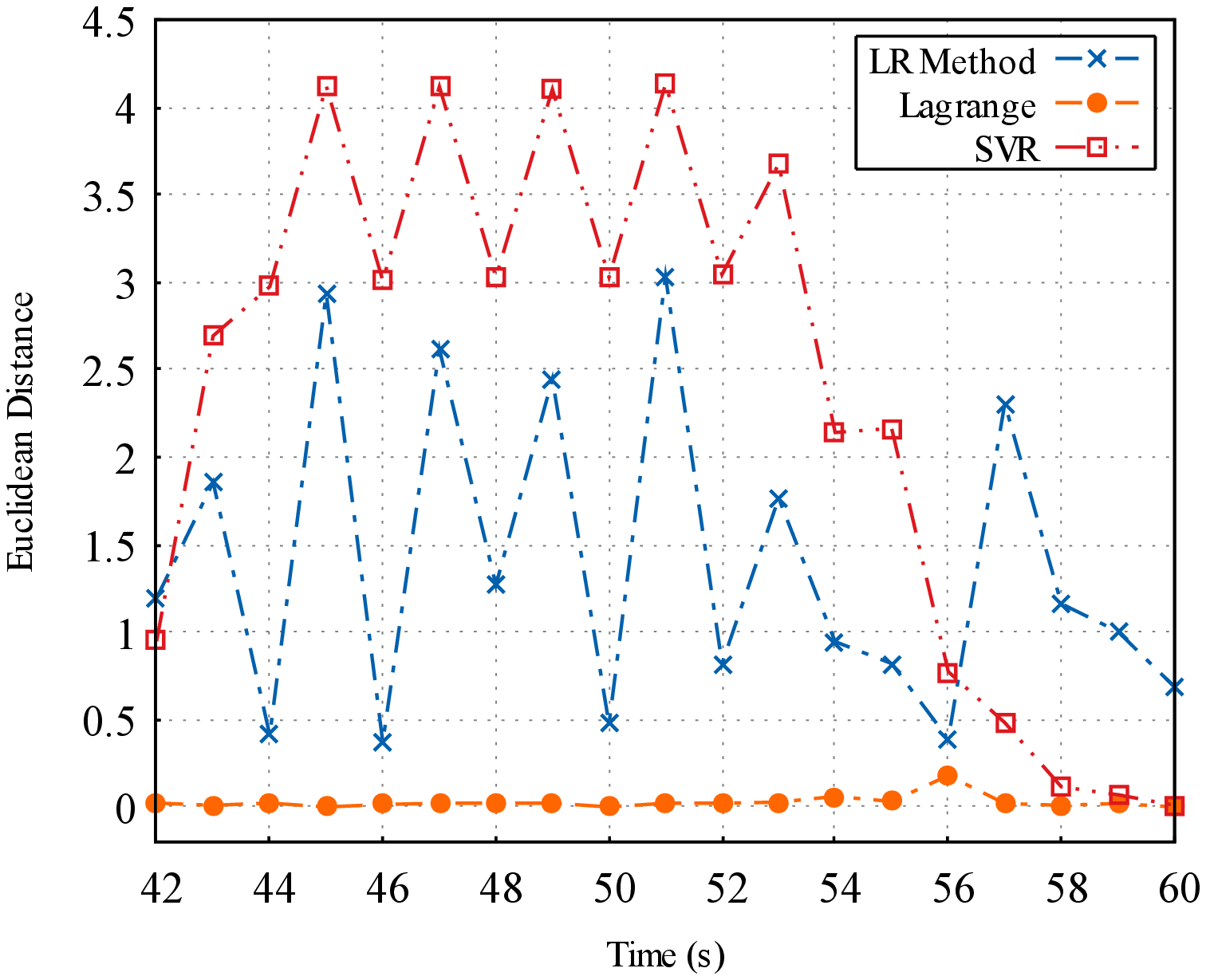}}
		\label{fig:performance_road}
	}
	\hfil
	\subfloat[Intersection Scenario]{
		{\includegraphics[width=0.31\linewidth]{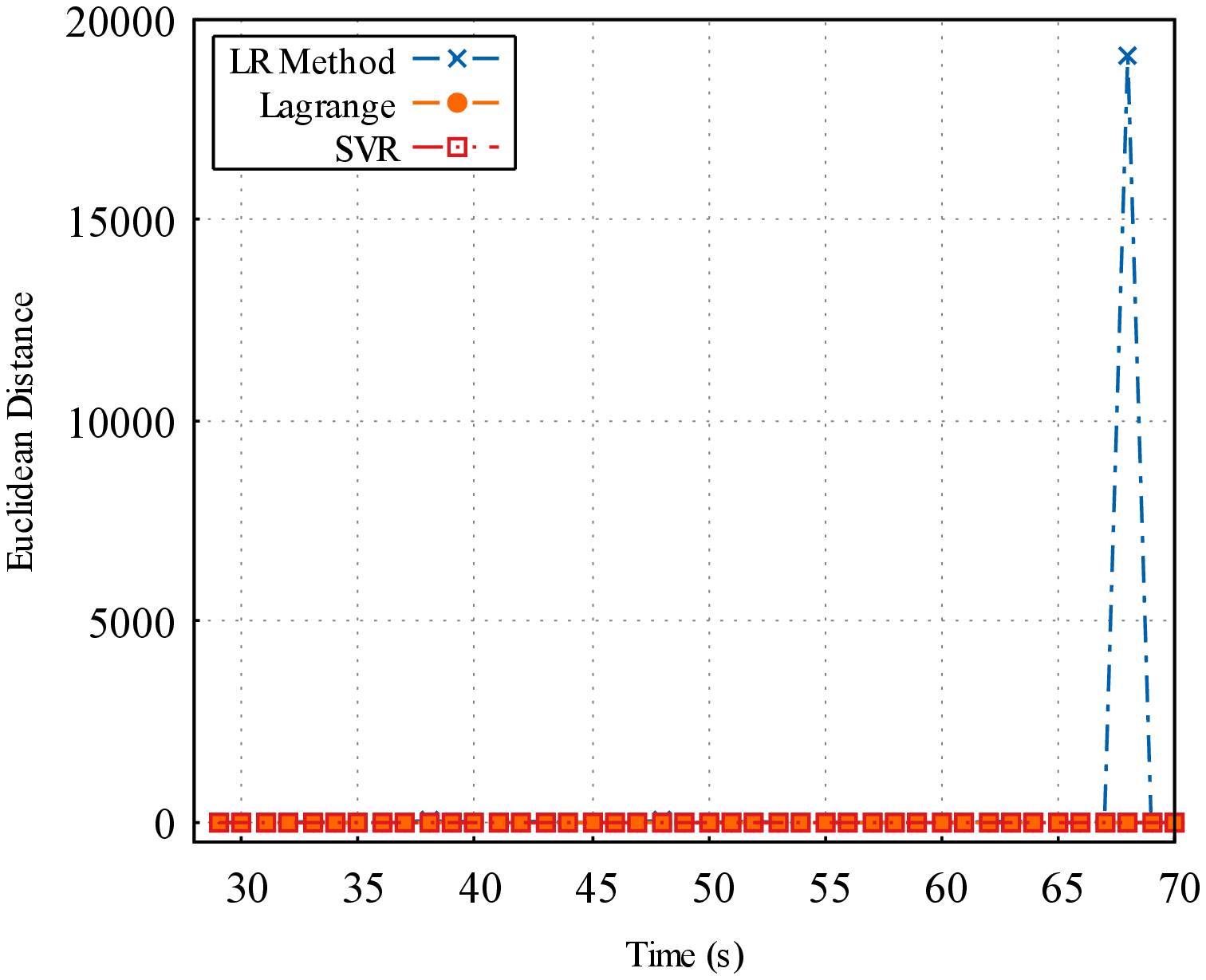}}
		\label{fig:performance_intersection}		
	}
	\hfil
	\subfloat[City Scenario]{
		{\includegraphics[width=0.31\linewidth]{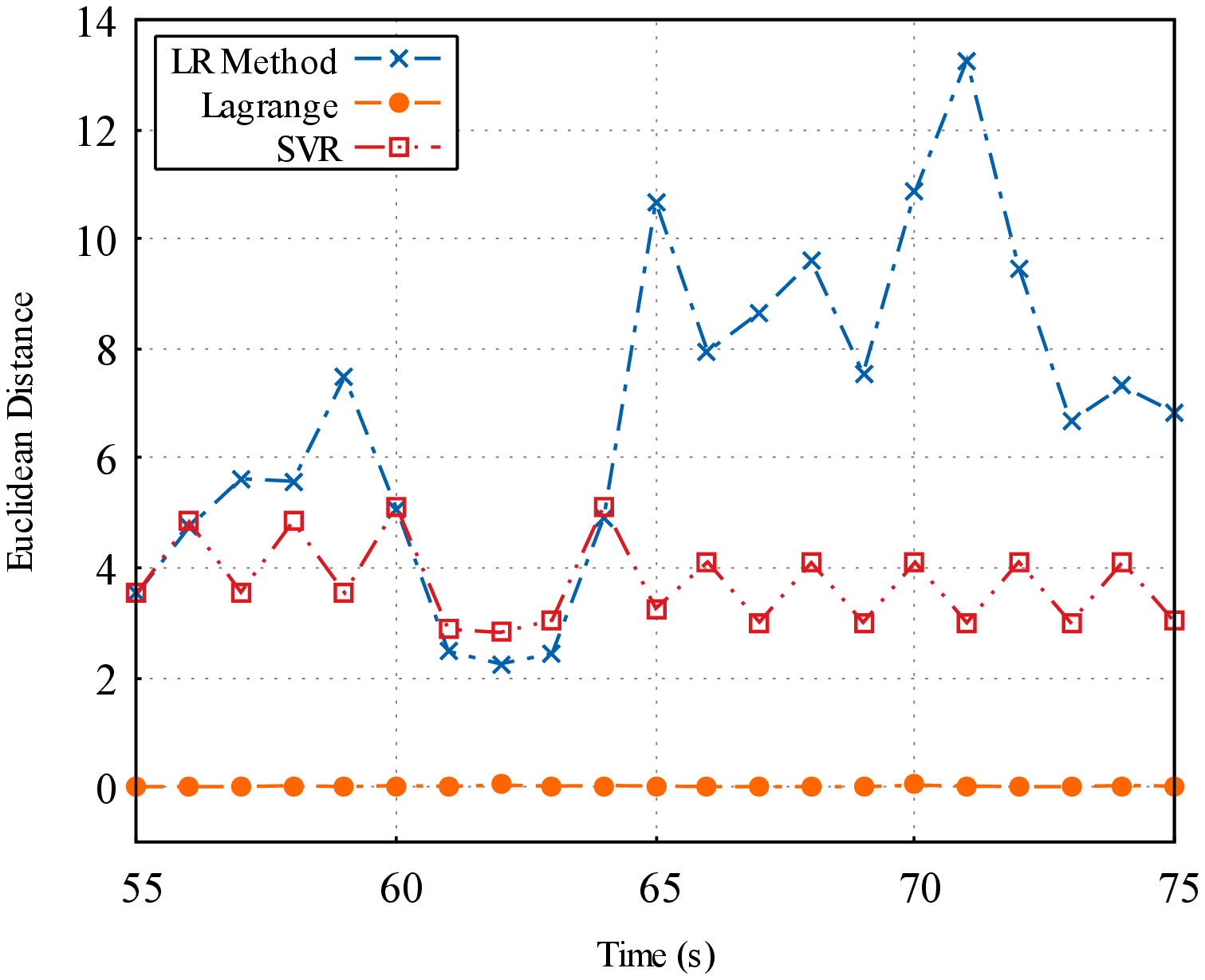}}
		\label{fig:performance_city}		
	}
	\caption{SVR, Lagrange \& LR Performance Analysis}
	\label{fig:performance_analysis}
\end{figure*}

In this subsection, we aim to present a comparative study between SVR and Lagrange methods. Therefore, we use the Euclidean distance to compute the distance between the real vehicular trajectory and the predicted trajectory using SVR and Lagrange methods.Formally, let $T_{r}=[(p_{1}, t_{1}), ...(p_{n}, t_{n})]$ be us consider as the real trajectory, where $t$ is an time instant and $p$ is the spatial position with $x_{r}$ and $y_{r}$ are the spatial coordinates. Let $T_{p}$ be is the predicted trajectory obtained with SVR or Lagrange method, where $x_{p}$ and $y_{p}$ are the spatial coordinates at a given instant $t_{i}$. Therefore, the similarity rate using the Euclidean distance can be expressed as follows:

\begin{equation}
dist = \sqrt[]{(x_{r} - x_{p})^2 + (y_{r} - y_{p})^2}
\end{equation}
where if the value of the distance $dist$ is equal to 0, then the predicted trajectory $T_{p}$  is similar to the real trajectory $T_{r}$ . \\
Figure \ref{fig:performance_road} reports the obtained similarity rates using LR, SVR and Lagrange methods in road scenario. According to the obtained results, we can see that the Lagrange method is better than SVR method, where the obtained distances are closer to zero, which indicate that predicted trajectory is almost similar to the real trajectory. In addition, the obtained distances with LR are very close to SVR distances.

Moreover, Figure \ref{fig:performance_intersection} shows the obtained distances in intersection scenario using LR, SVR and Lagrange. In fact, the obtained high distance value for SVR method is very close to 8 a the minimum distance value $dist$ is equal to 3,5 at $t=65, 67,$ and $69$ s. However, the Lagrange method gives better results compared to SVR and LR methods where its distance values are almost stable and close to zero for all instants from $29$ to $70$.

Finally, Figure \ref{fig:performance_city} reports the obtained distances values using three prediction methods on city scenario. In fact, Lagrange method remains almost stable for all instants,  whereas SVR method is almost to Lagrange where the minimum distance value $dist$ is considered as satisfactory, where it is equal to 2.8. However, LR is not better than SVR, where its distance values are not far than SVR and Lagrange methods.

To assess the proposed prediction approach SVR, Lagrange, and LR methods, we have used five performance metrics, including, mean square error (MSE), mean absolute error (MAE), root-mean-square error (RMSE), and mean absolute percentage error (MAPE). These evaluation criteria can be defined as : 

\begin{equation}
MSE = \frac{1}{n} \sum_{i=1}^{n} (P_{i} - \hat{P}_{i})^{2}
\end{equation}
\begin{equation}
MAE = \frac{1}{n} \sum_{i=1}^{n} |(P_{i} - \hat{P}_{i})|
\end{equation}
\begin{equation}
RMSE = (\frac{1}{n} \sum_{i=1}^{n} (P_{i} - \hat{P}_{i})^{2} )^{1/2}
\end{equation}
\begin{equation}
MAPE = \frac{100}{n} \sum_{i=1}^{n} \left| \dfrac{(P_{i} - \hat{P}_{i})}{P_{i}}  \right|
\end{equation}
where $n$ is the number of positions $P_{i}$ is the current position and $\hat{P}_{i}$ is the predicted position. Table \ref{tab:evaluation} reports the obtained values of MSE, MAE, RMSE and MAPE using LR, SVR, and Lagrange on the three scenarios city, intersection, and road.  According to obtained results, we can see that the Lagrange method is stable than SVR and LR. Moreover, SVR gives better results than LR in term of stability and precision.
 
\begin{table*}[!t]
	\centering
	\begin{tabular}{|l|c|r|r|c|r|r|c|r|r|c|r|r|} 
		\hline
		& \multicolumn{3}{c|}{MSE}& \multicolumn{3}{c|}{MAE}& \multicolumn{3}{c|}{RMSE}& \multicolumn{3}{c|}{MAPE} \\
		\hline
		Scenarios& LR & SVR & Lagrange & LR & SVR & Lagrange& LR & SVR & Lagrange& LR & SVR & Lagrange\\
		\hline
		City & 9.9514 & 7.2129  &2.8095e-04  & 3.1028&2.0055
		& 0.0138& 3.5410& 2.6857& 0.0168 &0.0921&0.0324&3.7166e-04\\
		\hline
		Road& 4.1247 & 3.8200  & 0.0011  &2.8751 & 1.2591 &0.0150 & 2.0158& 1.9545& 0.0331 &0.4216& 0.2233&0.0022 \\
		\hline
		Intersection &  11.2458&  10.5297 &  4.7143e-04 & 3.4857 &2.3181 & 0.0140& 4.9521&3.2450 & 0.0217 &0.1254&0.0571&4.8423e-04 \\
		\hline
	\end{tabular}
	\caption{Obtained values of evaluation criteria on LR, SVR and Lagrange}
	\label{tab:evaluation}
\end{table*}

\section{Conclusion}
This paper proposed new prediction methods of link failure using LR and SVR, allowing vehicles to control the state of link during the communications that may improve the quality of service in vehicular ad-hoc network (VANET). We studied the impact of SVR and LR in three different realistic mobility scenarios: road, intersection, and city communication case. Also, we compared the prediction results obtained using Lagrange, LR and SVR with real movement of vehicles in the three scenarios. The obtained results show that our methods is close to the reality movement of vehicle, which means that the decision obtained of link failure in future time using LR and SVR is more efficient.

\bibliographystyle{IEEEtran}
\bibliography{Laroui18GCRef}

\end{document}